\documentclass[%
 reprint,showkeys,
amsmath,amssymb,
 aps,
floatfix,
]{revtex4-2}
\usepackage{float}
\usepackage[caption=false]{subfig}
\usepackage[pdf]{pstricks}
\usepackage{mathtools}
\usepackage{braket}
\usepackage{graphicx}
\usepackage{dcolumn}
\usepackage{bm}
\usepackage{hyperref}
\begin{document}
\preprint{}
\title{Approximate Surface Tension of Phase Separating Interfaces using Residual Forces}

\author{Bal Krishan}
 \email{ph19051@iisermohali.ac.in}
\affiliation{%
 Department of Physical Sciences, Indian Institute of Science Education and Research, Mohali, Punjab, India}
\date{\today}
\begin{abstract}
 Extending methods that have been successful in determining the surface tension of equilibrium interfaces to active Brownian particle systems have resulted in negative and/or unusually large values of the interfacial tension, raising questions about the applicability of these frameworks to nonequilibrium interfaces. The absence of a general framework for understanding interfacial tension independently of equilibrium thermodynamics makes this problem particularly challenging. Here, I develop a new framework that provides an order-of-magnitude estimate of the interfacial tension using average force densities that are accessible even when the system is not in thermodynamic equilibrium. Applying this framework to active interfaces yields estimates that differ substantially in both sign and magnitude from previously reported results, highlighting the need to reassess existing approaches to interfacial tension in active matter.
\end{abstract}
\maketitle

\maketitle

\section{Introduction}
\indent A phase separated system possessing distinct high and low density regions naturally involves the formation of an interface region that separates the two regions with different densities. In a single component equilibrium fluid undergoing phase separation, density heterogeneity across the interface leads to inhomogeneous distribution of molecular forces in the interface region which are responsible for emergence of surface tension at the interface \cite{rowlinson1}. While the relationship between molecular forces, local stresses and surface tension is extensively studied for interfaces in passive systems \cite{rowlinson1,berry1971,grekov2021}, this is not the case for non-equilibrium systems such as those exhibiting an active interface. Even in passive systems, local stresses are not uniquely defined, though different formulations lead to consistent values of surface tension. But, it has proved to be more troublesome to define local stresses for active systems that can provide reliable values of surface tension. \\ 
\indent Swim pressure as introduced by Takatori \textit{et al.} \cite{takatori1} correctly represents the force on a boundary in active systems. It, however, fails to capture the subtle local behaviour required to evaluate the surface tension using the more commonly employed stress anisotropy route. Bialke \textit{et al.} \cite{bialke1} found that inclusion of swim pressure to determine the surface tension results in large negative values for the active interface formed by active Brownian particles (ABPs) interacting with repulsive only Weeks-Chandler-Anderson (WCA) potential in two dimensions. Since then, several attempts have been made to evaluate the surface tension of phase separating interface of ABPs. Using a similar approach, Patch \textit{et al.}\cite{patch1} also found large negative values of surface tension. Omar \textit{et al.} \cite{omar2} argued that due to its anisotropic nature, the swim stress is not a true stress and should more appropriately be understood as a force density. Using simulations of ABPs in three dimensions, they found that when swim stress is excluded, near zero values of surface tension are obtained and showed that inclusion of swim stress results in negative values of surface tension. Hermann \textit{et al.} \cite{sophie1} developed a theory based on chemical potential balance in the two phases across the interface and predicted that active interface has a positive surface tension. A rather unusual landscape is emerging from recent developments in phenomenological descriptions of active systems where the search for a single unifying interfacial tension characterizing surface behaviours is being abandoned for more context dependent surface tensions that can describe mechanical behaviour, capillary fluctuations and Ostwald ripening independently \cite{Cates2025,luke01}. See Li \textit{et al.} \cite{longfei2025} for an excellent overview of previous works, wherein a more generalized relation for surface tension of non-equilibrium interface is also derived which results in large positive values.\\
\indent Most previous works simulate ABPs in a periodic box, often elongated, where MIPS leads to formation of clusters. Due to elongated shape of the box, most clusters merge together to form one large cluster that spans the shorter length of the box leading to two interfaces parallel to the shorter length. Our approach, to analyze the active interface formed along longer side of box using confinement has previously been successful in obtaining insights into finite size scaling behaviour of the active interface, primarily due to access to a larger interface. We found that the active interface formed due to confinement exhibits three different growth mechanisms belonging to different universality classes during different stages of the growth \cite{bal2025}. In this work, I employ spatial coarse graining and extensive averaging to determine the average distribution of densities of forces experienced by the molecules for the system of confined ABPs and find that in the region of active interface, average self propulsion force acts antagonistic to average interparticle interaction force and is slightly larger leading to the formation of a stable interface. The resulting total force density is used to evaluate the amount of work required to move a particle from one bulk region to another, across the interface. A novel construction of the interface formation shows that this work done is an order-of-magnitude approximation to the surface tension of the interface, after accounting for dimensional corrections. Simulation of an equilibrium liquid-vapor interface made up spherical particles interacting with Lennard Jones potential are used to validate the proposed method and to help understand its limits.

\section{Simulation Details}
Simulation setup used in this work is exactly the same that I have used earlier to determine the finite size scaling and universality classes of the growing  active interface formed due to confinement \cite{bal2025}. Despite the resemblance, the present work addresses a different question and makes use of unrelated analysis. Our system consists of $N$ active Brownian particles confined between two rigidly held parallel walls moving in two dimensions whose dynamics is governed by the set of coupled overdamped Langevin equations,
\begin{eqnarray}
 \dot{\bm{r}_i} &=& D\beta(\bm{F}_i^{int}(r_{ij}) + \bm{F}_{i}^{wall}(r_{ij}) + \hat{v_i} v_p) + \sqrt{2D} \bm{\eta}_r \nonumber\\
 \dot{\theta_i} &=& \sqrt{2D_r} \eta_\theta \label{activelngvn}
\end{eqnarray}
Here, position and orientation coordinates of particles are represented by $\{\bm{r}_i,\theta_i\}$, $v_p$ denotes the magnitude of active force and its direction is given by the unit vector $\hat{v_i}=(\cos \theta_i,\sin \theta_i)$. Translational and rotational diffusion coefficients are written as $D$ and $D_r$ respectively which are taken to be related as $D_r = 3D/\sigma^2$ where $\sigma$ represents the diameter of the particles. Walls used in our simulations are made up of particles of same size as ABPs which are rigidly held at their positions. Inter-particle interaction force $\bm{F}_i^{int}(r_{ij})$ with $r_{ij} = |r_i - r_j|$ and the wall forces $\bm{F}_i^{wall}(r_{ij})$, are both obtained using the WCA potential\cite{weeks1},
\begin{equation}
U(r) =  \bigg\{ \begin{matrix} 4\epsilon\big[\big(\frac{\sigma}{r}\big)^{12} - \big(\frac{\sigma}{r}\big)^6 \big] + \epsilon \text{\hspace{1.1cm} if } r<2^{1/ 6}\sigma \\ 0 \text{\hspace{4cm} otherwise} \end{matrix}
\end{equation}
We take $D$, $\sigma$ and $\epsilon = 1/\beta$ to be one. For all practical purposes, the wall forces are interaction forces and from here on, are considered to be part of $\bm{F}_i^{int}$. Dimensionless Peclet number $Pe=v_p \sigma/D$ governs the magnitude of self propulsion force      $v_p$ and
$\{\eta_i\}$ are Gaussian white noise variables with
\begin{equation}
 \braket{\eta_i(t)} = 0  \text{\hspace{0.5cm}and\hspace{0.5cm}} \braket{\eta_i(t)\eta_j(t^\prime)} = \delta_{ij}\delta(t-t^\prime) \label{noisec}
\end{equation}
Above system of ABPs with periodic boundary conditions (in absence of walls) was first studied by Redner \textit{et al.} \cite{redner1} who determined the MIPS phase diagram of the system. We use a set of confining walls parallel to $x$-axis while retaining periodic boundaries in the other direction to obtain access to the active interface. Presence of confinement leads to preferential sorting of particle orientations with the particles that point upwards (downwards) getting accumulated at the upper (lower) confining walls.\\
\indent To determine the spatial distribution of molecular forces, we carried out simulations for $N=100172$ ABPs confined between parallel walls in a box of size $800 \times 200$ for different $Pe$ values. Time step is taken to be $1.0\times 10^{-5}$ for $Pe \ge 100$ and $2.0\times 10^{-5}$ for smaller $Pe$ values and after 100 units of equilibration time, data is collected for the next 100 units of times at the rate of 50 uniformly spaced snapshots for every unit of time elapsed. \\
\indent I also simulate a passive liquid-vapor interface of spherical particles interacting with Lennard-Jones (LJ) potential with a cut-off at $2.5\sigma$ where $\sigma$ represents the particle diameter,
\begin{equation}
U(r) =  \bigg\{ \begin{matrix} 4\epsilon\big[\big(\frac{\sigma}{r}\big)^{12} - \big(\frac{\sigma}{r}\big)^6 \big]  \text{\hspace{1.1cm} if } r<2.5\sigma \\ 0 \text{\hspace{3.5cm} otherwise} \end{matrix}
\end{equation}
with $\epsilon$ and $\sigma$ equal to one. Similar to the case of ABPs, a set of confining walls parallel to the z-axis are used which comprise of three layers of compactly arranged particles of similar size as fluid particles stacked on top of each other and are held fixed at their position. The attraction strength between the wall and fluid particles, $\epsilon_{\text{wall-fluid}}$ is taken to be 2. This confined geometry is chosen to avoid the complexity associated with the drift of the liquid region in the absence of walls. This system comprises of 95256 fluid particles in a box of size $160\times 16 \times 80$ and 16422 wall particles that extends outside $|z|\ge 40.0$ on both sides. In this case, time step is taken to be $5\times 10^{-4}$. After 1000 units of equilibration time, data is collected for the next 1000 units at the rate of 10 snapshots per unit of elapsed time.

\section{Method and Results}
\begin{figure}[!t]
  \centering
  \includegraphics[scale=.67]{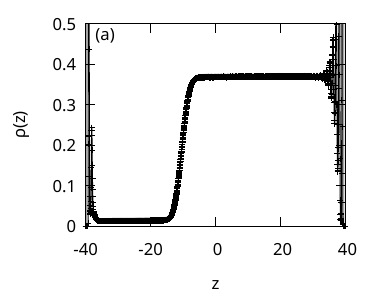}
  \includegraphics[scale=.67]{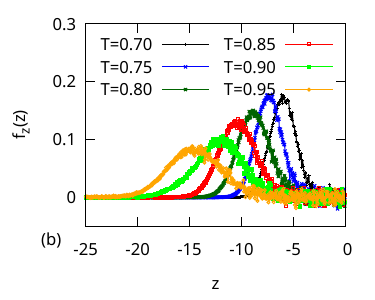}
	\caption{ For a phase separated system of spherical particles interacting via the Lennard Jones potential in presence of attractive walls: (a) average density distribution as a function of the z-coordinate at temperature $T=0.85$ showing the presence of a liquid vapor interface. (b) Average force density in the normal ($z$) direction for different $T$ values.
}
  \label{fig:lj}
\end{figure}

\indent I first outline a method for approximating the surface tension of a flat interface based on the definition of surface tension as the amount of work required to create an interface of unit area i.e.
\begin{align}
 \Delta W = \gamma dA \label{thermodfn}
\end{align} 
where $\Delta W$ is the amount of work required to create an interface of surface area $dA$ where none existed before and $\gamma$ represents the surface tension of the interface. Consider breaking down the process of creation of the interface as follows: starting from a homogeneous system, we move particles in an infinitesimally small volume $dV \sim \sigma^3$ (which is $\ll V$), $\sigma$ being the diameter of particles, from one region to another, and keep on repeating the process until the desired density heterogeneity is achieved. Let $\Delta W_{i-1}$ represent the amount of work done during $i^{th}$ step and let $n$ steps be required to reach the steady state density distribution. We can obtain the work required to create the interface by adding the work required during each of these $n$ steps,
\begin{align}
 \Delta W = \Delta W_0 + \Delta W_1 + ... + \Delta W_{n-1} \label{worksum}
\end{align} 
Because we start from a homogeneous system with uniform density, $\Delta W_0 = 0$. Also, as work done during each consecutive step results in an increase in anisotropy of forces at the interface, $\Delta W_{i} > \Delta W_{i-1}$ i.e. successive terms in above expression are monotonically increasing. At present, there is no way to exactly determine the terms on the right side of Eq. \eqref{worksum}. We can however, determine the $\Delta W_n$ term with the help of average force densities obtained during the steady state,
\begin{align}
 \Delta W_n =  \int^{z_{L}}_{z_V} dz [f_z^{}(z) \sigma^3 ] \label{work2}
\end{align}
where $z_L$ and $z_V$ are two points that lie inside the bulk of liquid and vapor phase respectively, $z$ being the direction normal to the interface and $f_z(z)$ represents the average (total) force density in the direction normal to the interface, during the steady state. Due to symmetry, other components of the (total) force density in the direction parallel to the interface averages to zero. The volume element $dV$ is infinitesimal as required by the continuum mechanics and is taken to accommodate a single particle $dV = \sigma^3$, so as to minimally affect the existing particle and force densities during each step. Since $\Delta W_{i} > \Delta W_{i-1}$, and we can not satisfactorily determine $\Delta W_{i}$ except when $i=n$, and it seems only reasonable to take the work required to create the interface with $\Delta W_{i=n}$ as a first approximation i.e. we take $\Delta W \approx \Delta W_n$. Comparing $\Delta W_n$ from Eq. \eqref{work2} with $\Delta W$ in Eq. \eqref{thermodfn}, we get
\begin{align}
	\gamma_{\text{approx}} = \sigma\int^{z_{L}}_{z_V} dz f_z^{}(z) \label{surface_tension2}
\end{align}
where we have used $dV/dA = \sigma$. Presence of microscopic length scale $\sigma$, in the above equation ensures that approximation surface tension $\gamma_{\text{approx}}$, has the correct dimension of force per unit length and is a direct consequence of the fact that the emergence of tension at an interface is the macroscopic manifestation of a phenomena that essentially plays out at microscopic scales. While the exact surface tension is not directly proportional to the size of liquid molecules, it is observed to increase with size of fluid particles at fixed temperature. \\
\indent Even though the approximation considered above can never yield the exact work required to create the interface, we expect it to provide a value of surface tension that has the same order as the exact surface tension, primarily due to monotonically increasing nature of successive work terms in Eq. \eqref{worksum}. We first test this method for the equilibrium Lennard Jones liquid vapor interface for which exact surface tension values are known with reliable accuracy. In Fig. \ref{fig:lj}(a), a specimen density profile for this interface is shown for temperature $T=0.85$ with the simulation setup discussed earlier. Force density profiles in the direction normal to the interface are shown in Fig. \ref{fig:lj}(b) for different $T$ values, while the other components vanish upon averaging. To determine the approximate surface tension given by Eq. \eqref{surface_tension2}, I integrate the force density profiles shown in Fig. \ref{fig:lj}(b) and compare them with the surface tension values obtained by the stress anisotropy integral, both of which are shown in Table \ref{tab:one}. Two observation in Table \ref{tab:one} are worth noting: i) the reliability of the approximate method drops significantly as the critical point ($T \sim 1.05$) \cite{vrabec2006} is approached, and ii) in comparison to the exact method, the approximate method is quite insensitive to the temperature. While the first observation is useful for practical purposes, the second one provides an opportunity to better understand the approximate method itself. \\
\begin{table}[!b]
\begin{center}
	\caption{For the Lennard Jones liquid vapor interface formed due to confinement, approximate surface tension values using method outlined here and exact surface tension using the stress anisotropy method are shown. Exact values are in excellent agreement with those without the confinement \cite{simon2018} indicating the presence of confinement has no significant impact on the surface properties. As the critical point is approached, the performance of the approximate method drops drastically.}  
\label{tab:one}
  \begin{ruledtabular}
\begin{tabular}{ l  c  c  r }
	$T$ \hspace{0.8cm}  & \hspace{0.8cm} $\gamma_{\text{approx}}$ \hspace{0.8cm} & \hspace{0.8cm} $\gamma$ \hspace{0.8cm} & \hspace{0.8cm} Error \\
	 & (Eq. \eqref{surface_tension2}) & (Exact) & (in \%age)\\
\hline
0.65 & 0.52 & 0.68 & 23.5\\
0.70 & 0.54 & 0.59 & 8.5\\
0.75 & 0.56 & 0.50 & 12.0\\
0.80 & 0.57 & 0.40 & 42.5\\
0.85 & 0.57 & 0.32 & 78.1\\
0.90 & 0.56 & 0.24 & 133.3\\
0.95 & 0.54 & 0.15 & 260.0\\
\end{tabular}
  \end{ruledtabular}
\end{center}
\end{table}
\indent For a planar interface at mechanical equilibrium, the normal component of the pressure tensor must be spatially invariant ($dP_{zz}(z)/dz = 0$). Decomposing this tensor into kinetic and configurational parts leads to,
\begin{align}
	\frac{d}{dz}[\rho(z)k_B T + P_{zz}^{\text{conf}}(z)] = 0
\end{align}
where $\rho(z)$ represents the density profile and $P_{zz}^{\text{conf}}$ is the configurational contribution to the pressure tensor. Because the local intermolecular force density isolates the gradient of the configurational stress ($f_z(z) = - dP_{zz}^{\text{conf}}/dz$), the condition of the hydrostatic equilibrium simplifies to $f(z) = k_B T \frac{d\rho(z)}{dz}$. Integrating this expression across the interface, we get
\begin{align}
	\int_{z_V} ^{z_L} f(z) dz = k_B T (\rho_L - \rho_V) \label{hydroeqm2}
\end{align}
where $\rho_L$ and $\rho_V$ are densities of the liquid and vapor phases respectively. Consequently, the approximate method [Eq. \eqref{surface_tension2}] evaluates a product related to bulk coexistence properties. The observed temperature insensitivity ($\gamma_{\text{approx}} \sim 0.55$) in the temperature range $0.7\le T \le 0.95$ is a direct mathematical consequence of the product on right side of Eq. \eqref{hydroeqm2}, where the linear increase in temperature is being compensated by the evaporation driven drop in the coexisting density differences. Nevertheless, as shown in Table \ref{tab:one}, the approximate method remains a useful tool to obtain order-of-magnitude estimate for the surface tension of a flat interface, far away from the critical point and is especially useful for non-equilibrium situations such as the active interface where extensions of equilibrium based approaches have repeatedly failed to yield reliable estimates.\\
 \begin{figure}[!t]
  \centering
  \includegraphics[scale=0.67]{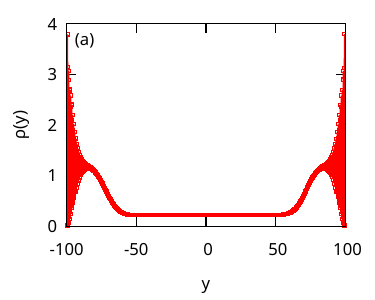}
  \includegraphics[scale=0.67]{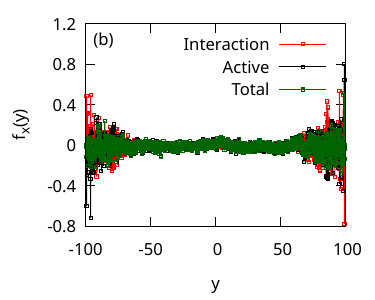}
  \includegraphics[scale=0.67]{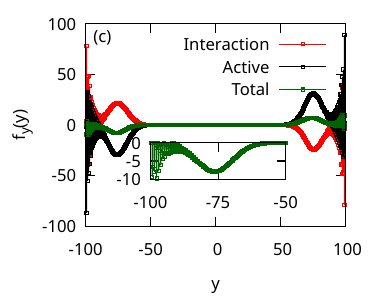}
  \includegraphics[scale=0.67]{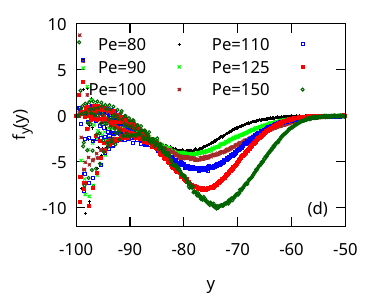}
	 \caption{For a system of confined ABPs with $Pe=125$, (a) average particle density and average force densities for both interaction and active forces in the direction, (b) parallel to the interface, and (c) normal to the interface, as a function of the $y$-coordinate. Average force densities parallel to the interface are negligible whereas those normal to the interface exhibit extremum values near the mean interface position. Inset in (c) zooms in on the region of lower interface and shows the total force density as a function of the y-coordinate. (d) Normal force density profiles at the active interface for different activity ($Pe$) values. 
}
  \label{fig:deltayforces}
 \end{figure}
\indent 
\indent For the active interface, variations in particle density normal to the interface are shown in Fig. \ref{fig:deltayforces} (a) for $Pe=125$, where it can be observed that in the high density region, the density exhibits oscillating behaviour. These oscillations indicate the emergence of strong solid-like density modulations near the confining substrate which decay as we move away from the confinement. As reported in \cite{bal2025}, average interface position for $Pe=125$ is $\sim 27\sigma$ away from the confining substrate and is expected to have little or no effect of the induced particle order near the confinement. To avoid confusion with units of surface tension for systems in two and three dimensions, we think of the system of circular disks as a system of spherical particles confined within an extent $L_z = 1$. In the remaining, we deal with this quasi two-dimensional interpretation and the force densities are to be treated as force per unit volume even when division by $L_z$ is not explicitly written. 
The simulation box is divided into tiny sub-boxes along y-axis that span the entire length along x-axis. In each sub-box, the swim and interaction forces on all particles are added together and divided by the area of the sub-box to obtain the average density of force $\bm{f}^{}$ inside the sub-box i.e.,
\begin{align}
\bm{f}^{}(y) =  \bm{f}^{Act}(y) + \bm{f}^{Int}(y) \label{forcesum}
\end{align}
Here, $\bm{f}^{Act}$ and $\bm{f}^{Int}$ are respectively the average active and interaction force per unit area in the strip at $y$,
\begin{align}
\bm{f}^{Act}(y) = \frac{1}{L_x\Delta y}\sum_i \bm{F}_i^{Act} \nonumber \\
\bm{f}^{Int}(y) = \frac{1}{L_x\Delta y}\sum_i \bm{F}_i^{Int} 
\end{align}
where, $\bm{F}_i^{Act}$ and $\bm{F}_i^{Int}$ are interaction and active forces on $i^{th}$ particle respectively and the summation extends over all particles in the strip at $y$. It is worth noting that frictional forces are ignored in the order-of-magnitude estimation presented here, as they are isotropically distributed in the lower density phase due to randomized particle motion and are negligible in the high density region where particle velocities are very small. As expected due to symmetry, parallel components of both forces, $f_{x}^{Act}(y)$ and $f_{x}^{Int}(y)$ are found to be negligible as shown in Fig. \ref{fig:deltayforces} (b).  Normal to the interface, average density for both the forces is found to exhibit extremum values near the mean interface position, which is shown in Fig. \ref{fig:deltayforces} (c). Total force density in the normal direction is also shown in Fig. \ref{fig:deltayforces} (c) and inset therein, where a zoom in on the lower interface region is shown. It can be observed that the active force density is slightly larger and opposite to the interaction force density which leads to the formation of a stable interface. Total force density for different $Pe$ values is shown in Fig. \ref{fig:deltayforces}(d).\\ 
\indent Oscillating behaviour of force density due to density modulations near the confinement seen in Fig. \ref{fig:deltayforces}(d) complicates a direct evaluation of the integral in Eq. \eqref{surface_tension2}. To overcome this difficulty, force density profile is assumed to be symmetric about the interface which is also supported by the observation on the Lennard Jones liquid vapor interface shown in Fig. \ref{fig:lj}(b). Symmetric half of the integral in Eq. \eqref{surface_tension2} is than evaluated by replacing the upper limit by the position of the extremum of the force density which is than multiplied by two to obtain $\gamma_{\text{approx}}$ which are shown in Table \ref{tab:two}. The resulting approximate surface tension values, are found to be order $\sim100$ (in units of $\epsilon/\sigma^2$), in clear contrast to the near-zero, negative or large positive values reported from stress anisotropy and related frameworks in earlier works \cite{bialke1,patch1,omar2,longfei2025}. The approximated tension shows a systematic dependence on the magnitude of active force ($Pe$), suggesting a progressive strengthening of interfacial cohesion with increasing activity.
\begin{table}[!t]
\begin{center}
\caption{Order-of-magnitude estimates for the surface tension of the active interface obtained using Eq. \eqref{surface_tension2}.}
\label{tab:two}
  \begin{ruledtabular}
\begin{tabular}{ c  c  }
	$Pe$  & $\gamma_{\text{approx}}$ \\
 (Activity)  &  (Surface tension)  \\
\hline
80  &  70.40 \\
90  &  85.82 \\
100 &  103.81 \\ 
110 &  109.57 \\ 
125 &  143.98 \\
150 &  175.31 \\
\end{tabular}
\end{ruledtabular}
\end{center}
\end{table}

\section{Discussion}
I have determined the spatial distributions of the densities of molecular forces experienced by the particles in a system of confined ABPs. Despite rapid fluctuations in forces experienced by individual particles, average total force density is non-zero only in the direction normal to the interface. Within the interfacial region, the active (swim) force density is found to be slightly larger than the interaction force density, which leads to a stable interface. The antagonistic interplay between these two forces determines the work required to move particles from one bulk phase to the other across the interface. After accounting for dimensional corrections, this work is shown to be of the same order of magnitude as the interfacial surface tension. The approximate surface tension values obtained for active interfaces are positive, moderately large, and increase progressively with activity. These results are in contrast with previous approaches that evaluate active and interaction contributions to the surface tension independently or rely on other equilibrium based frameworks.\\
\indent The approach developed in this work emphasizes the importance of considering the collective effect of competing forces. Since the method relies on spatially averaged force densities and approaches the macroscopic limit through extensive averaging, its predictions are amenable to experimental validation. Although the present analysis focuses on the two dominant forces in ABPs, further refinements are needed to account for additional contributions, including frictional forces and alignment interactions in chiral or Vicsek-like active particles, which may alter the force-density distributions and the resulting interfacial tension. More broadly, the discrepancies between the present estimates and previous results underscore the need for a more consistent understanding of interfacial tension in nonequilibrium systems. Despite the approximations inherent in the present framework, these results provide a physically motivated reference for assessing existing and future approaches to interfacial tension in nonequilibrium systems and are expected to motivate the development of more refined frameworks for determining surface tension and other interfacial properties.

\section*{Acknowledgement}
I express my sincere gratitude to my PhD advisor, Abhishek Chaudhuri, for his mentorship and valuable guidance throughtout this work. For compute resources, I acknowledge HPC Facility at IISER Mohali.

\bibliographystyle{apsrev4-1}
\bibliography{references}

\end{document}